\documentclass[twocolumn,floatfix,showpacs]{revtex4}
\usepackage{amsmath,amsfonts,amssymb}
\usepackage[dvips]{graphicx}
\usepackage{ulem}
\usepackage{color}
\newcommand{\nc}{\newcommand}
\nc{\rnc}{\renewcommand}
\nc{\nn}{\nonumber}
\nc{\db}{\displaybreak[0]\\}
\nc{\ds}{\displaystyle}
\rnc{\[}{\left[}
\rnc{\]}{\right]}
\rnc{\(}{\left(}
\rnc{\)}{\right)}
\nc{\lt}{\left\{}
\nc{\rt}{\right\}}
\nc{\om}{\omega}
\nc{\sg}{\sigma}
\nc{\lam}{\lambda}
\rnc{\a}{\alpha}
\nc{\bra}{\langle}
\nc{\ket}{\rangle}
\rnc{\i}{{\rm i}}
\rnc{\d}{{\rm d}}
\nc\nt{\tilde{n}}
\nc\gt{\tilde{g}}
\nc\Gt{\tilde{G}}
\nc{\D}{\Delta}

\nc\vac{|0\ket}
\nc\sn{|S_N\ket}
\nc\tn{|T_N\ket}
\nc\tm{|T_m\ket}
\nc\uq{U_q(\mathfrak{sl}_2)}
\nc\mm{\mathcal{M}}

\nc{\red}{\textcolor{red}}

\begin{document}
\title{
Relaxation dynamics of closed diffusive systems with infinitesimal Langmuir kinetics
}
\author{Jun Sato}
\author{Katsuhiro Nishinari}
\address{
Research Center for Advanced Science and Technology, University of Tokyo, \\\it 
4-6-1 Komaba, Meguro-ku, Tokyo 153-8904, Japan
}
\date{\today}
\begin{abstract}
We consider the asymmetric simple exclusion process with Langmuir kinetics 
in the closed boundary condition. 
We analytically obtain the exact stationary state 
and a series of excited states of the system 
in the limit where Langmuir kinetics is infinitesimally small. 
Based on this result, we propose an analytical formula for the time evolution 
of physical quantities of the system.
\end{abstract}
\pacs{05.10.Gg}
\maketitle
\section{Introduction}
Recently, one-dimensional driven-diffusive systems have attracted much interest 
in the context of nonequilibrium statistical physics. 
Among them, 
the asymmetric simple exclusion process (ASEP) is one of the most fundamental and exactly solvable models 
describing the nonequilibrium transport phenomena \cite{MGPB, DEHP, KSKS, Sasamoto, Schadschneider, SCN, Schutz2000}. 
It has a wide range of applications such as in 
biology \cite{MGPB} and 
pedestrian and traffic flow \cite{Schadschneider, SCN}.

The ASEP is a continuous-time Markov process 
describing the asymmetric diffusion of particles with an exclusion principle on a one-dimensional lattice. 
In a time interval $\d t$, 
a particle hops to the right (left) site with probability $p\d t$ $(q\d t)$ if it is vacant. 
The Markov matrix describing the time evolution of the ASEP 
can be exactly diagonalized by the Bethe ansatz \cite{GM} 
and the exact stationary state can be constructed by the matrix product ansatz \cite{BE}. 
These methods offer exact derivations of the interesting phenomena 
such as boundary-induced phase transitions \cite{BE, BECE, Krug}. 
Moreover, it is shown that the current fluctuation belongs to the Kardar-Parisi-Zhang universality class \cite{kpz} 
using the random-matrix theory \cite{random}. 

The ASEP with Langmuir kinetics (ASEP-LK) 
describes an attachment and detachment of particles 
as well as the exclusive hopping process \cite{PFF}. 
In this system, particles go in and out anywhere on the lattice, 
whereas in the usual open boundary system, 
particle exchange with the outer system occurs only at the end of the system. 
Steady-state properties of this model 
have been studied in detail 
by use of mean-field theory and Monte Carlo simulations \cite{EJS, ISN1, ISN2}. 
These studies reveal interesting phenomena such as 
the coexistence of high- and low-density phases separated by the shock wave in the density profile. 
The exact stationary state is constructed in the case of periodic boundaries \cite{EN}. 
More recently, 
the exact time evolution of correlation functions in the ASEP-LK with periodic lattice was obtained in \cite{SN}. 

On the other hand, it was shown that the ASEP with a closed boundary has $U_q(\mathfrak{sl}_2)$ symmetry \cite{SS1994}. 
This means that the generators of $U_q(\mathfrak{sl}_2)$ algebra commute with the Markov matrix 
describing the time evolution of the system. 
Using this fact, the $N$-particle steady state $\sn$ can be constructed 
by applying the creation operator $F$, $N$ times on the vacuum $\vac$: 
$\sn\propto F^N\vac$. 

In this paper we consider the closed ASEP-LK described in Fig. \ref{model}. 
The system loses its $\uq$ symmetry due to the Langmuir kinetics (LK). 
The stationary state can no longer be written in a closed form, 
unlike in the case of a periodic boundary with LK. 
However, we find that the stationary state has a closed form in the case of infinitesimally small LK. 
Moreover, we obtain a series of low-lying excitations in this limit. 
Inserting these states into the time-dependent expectation value, 
we obtain a formula for the time evolution of the physical quantity of the system 
starting from the vacuum as an initial state.

\section{Model and notations}
\begin{figure}
\includegraphics[width=0.8\columnwidth]{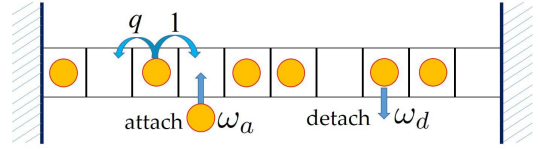}
\caption{
The ASEP-LK with a closed boundary. 
}
\label{model}
\end{figure}

We consider the ASEP-LK with a closed boundary, 
which is schematically shown in Fig. \ref{model}. 
A particle hops to the right (left) site with a rate $p=1$ $(q)$ if it is vacant. 
Particles are attached on a site with rate $\om_a$ if the site is vacant 
and detached with rate $\om_d$ if it is occupied. 

Hereafter we impose a closed boundary condition. 
We denote the number of lattice sites by $L$. 
We associate a Boolean variable $\tau_n$ to every site $n$ 
to represent whether a particle is present $(\tau_n=1)$ or not $(\tau_n=0)$. 
Let $|\tau_n=0\ket$ and $|\tau_n=1\ket$ denote the standard basis vectors in this order for the vector space $\mathbb{C}^2$. 
We consider the $L$-fold tensor product of this basis 
$|\tau_1\tau_2\cdots\tau_L\ket:=|\tau_1\ket\otimes|\tau_2\ket\otimes\cdots\otimes|\tau_L\ket$, 
the dimension of which is $2^L$. 
Then we can write a state of the system at time $t$ in a vector form 
with each element being a probability distribution 
\begin{align}
|P(t)\ket=\sum_{\tau_1,\cdots,\tau_L}
P(\tau_1,\cdots,\tau_L|t)|\tau_1,\cdots,\tau_L\ket. 
\end{align}
The time evolution of this state is described by the master equation
\begin{align}
\frac{\d}{\d t}|P(t)\ket
=\mm|P(t)\ket, 
\label{master}
\end{align}
where the Markov matrix $\mm$ is given by
\begin{align}
&\mm=\sum_{n=1}^{L-1}\mm_{n,n+1}+\sum_{n=1}^{L}h_n, 
\nn\\&
\mm_{n,n+1}=
\begin{pmatrix}
0& 0& 0&0\\
0&-q& 1&0\\
0& q&-1&0\\
0& 0& 0&0
\end{pmatrix}
_{\!\!n,n+1}, 
\nn\\&
h_n=
\begin{pmatrix}
-\om_a& \om_d\\
 \om_a&-\om_d
\end{pmatrix}
_{\!\!n}. 
\end{align}
The subscripts at the bottom right of the matrix represent the vector space on which the matrix is acting. 
The matrices operate as an identity elsewhere. 
The LK term $h_n$ can be regarded as an off-diagonal magnetic field 
in the language of quantum spin chains, 
which induces a nonconservation of the number of particles in the system.  

The stationary state $|S\ket$ belongs to the eigenvector of the Markov matrix 
associated with a zero eigenvalue:
\begin{align}
\mm|S\ket=0. 
\end{align}
The existence of the physically meaningful (all the elements are non-negative real numbers) 
unique stationary state is guaranteed by the Perron-Frobenius theorem for stochastic matrices. 
\section{Closed ASEP without Langmuir kinetics}
\subsection{$\uq$ Symmetry}
The ASEP with a closed boundary condition has $U_q(\mathfrak{sl}_2)$ symmetry \cite{SS1994}. 
In the following we will fix the value of the parameter $q$ in the interval $0<q<1$. 
The quantum group $\uq$ is 
the algebra generated by 
$e,f,k$ and $k^{-1}$
with the defining relations
\begin{align}
&[e,f]=\frac{k-k^{-1}}{q^{\frac12}-q^{-\frac12}}, 
\hspace{3mm}
kk^{-1}=k^{-1}k=1, 
\nn\\&
kek^{-1}=qe, 
\hspace{3mm}
kfk^{-1}=q^{-1}f. 
\label{def_uq}
\end{align}
The formal substitution $k^{\pm1}=q^{\pm h/2}$ and the limit $q\to1$ recover 
the usual commutation relations of the Lie algebra $\mathfrak{sl}_2$, 
\begin{align}
[e,f]=h, \quad
[h,e]=2e, \quad
[h,f]=-2f. 
\end{align}
The universal enveloping algebra $U(\mathfrak{sl}_2)$ is 
the noncommutative polynomial ring over $\mathbb{C}$ 
with the variables $e, f$ and $h$ divided by the ideal generated by
$ef-fe-h$, $he-eh-2e$ and $hf-fh+2f$. 
The quantum group $\uq$ is considered to be a $q$-deformation of $U(\mathfrak{sl}_2)$. 

In our model, each site has two states, namely, it is empty or occupied. 
Thus we adopt a two-dimensional representation of $\uq$ given explicitly by
\begin{align}
&e=
\begin{pmatrix}
0& 1\\
0& 0
\end{pmatrix}
, \quad
f=
\begin{pmatrix}
0& 0\\
1& 0
\end{pmatrix},
\nn\\&
k=
\begin{pmatrix}
q^{\frac12}& 0 \\
0 & q^{-\frac12}
\end{pmatrix}
, \hspace{3mm}
k^{-1}=
\begin{pmatrix}
q^{-\frac12}& 0 \\
0 & q^{\frac12}
\end{pmatrix}.
\end{align}
The defining relations \eqref{def_uq} are easily checked by the direct calculation of $2\times2$ matrices. 
The operators $e$ and $f$ are considered as annihilation and creation operators, respectively. 
We slightly change the normalizations of $k^{\pm1}$ by introducing $g$ and $\gt$ as 
\begin{align}
g=q^{\frac12}k^{-1}=
\begin{pmatrix}
1& 0\\
0& q
\end{pmatrix}
, \hspace{3mm}
\gt=q^{\frac12}k=
\begin{pmatrix}
q& 0\\
0& 1
\end{pmatrix}, 
\end{align}
which yields defining relations 
\begin{align}
&[e,f]=\frac{g-\gt}{1-q}, 
\hspace{2mm}
g\gt=\gt g=q, 
\hspace{2mm}
\gt eg=q^2e, 
\hspace{2mm}
\gt fg=f. 
\end{align}
In order to construct a tensor product representation acting on the total lattice of the system, 
we define the co-product structure as
\begin{align}
&\D(e)=e\otimes1+g\otimes e, \quad
 \D(f)=f\otimes\gt+1\otimes f, \nn\\
&\D(g)=g\otimes g, \quad
\D(\gt)=\gt\otimes \gt. 
\end{align}
Due to the co-associativity 
$(\D\otimes\text{id})\circ\D=(\text{id}\otimes\D)\circ\D$, 
we can define the operators acting on a total lattice
\begin{align}
&E:=\D^{(L-1)}(e)=e_1+g_1e_2+\cdots+g_1\cdots g_{L-1}e_L, \nn\\
&F:=\D^{(L-1)}(f)=f_1\gt_2\cdots\gt_{L}+\cdots+f_{L-1}\gt_L+f_L, \nn\\
&G:=\D^{(L-1)}(g)=g_1\cdots g_{L}, \nn\\
&\Gt:=\D^{(L-1)}(\gt)=\gt_1\cdots\gt_{L}. 
\end{align}
Here we adopt the abbreviated notation such that, for example, 
$x_2y_5:=1\otimes x \otimes 1 \otimes 1 \otimes y \otimes 1$ 
in the case of $L=6$. 
Since the co-product $\D$ conserves the defining relations 
\begin{align}
&\D\([e,f]\)=\D\(\frac{g-\gt}{1-q}\), \quad
\D\(g\gt\)=\D\(\gt g\)=q, 
\nn\\&
\D\(\gt eg\)=q^2\D\(e\), 
\quad
\D\(\gt fg\)=\D\(f\), 
\end{align}
the total operators $E, F, G$ and $\Gt$ satisfy the same relations 
\begin{align}
&[E,F]=\frac{G-\Gt}{1-q}, \quad
G\Gt=\Gt G=q, 
\nn\\&
\Gt EG=q^2E, 
\hspace{3mm}
\Gt FG=F. 
\end{align}
We can show that the ASEP-LK with a closed boundary has the $U_q(\mathfrak{sl}_2)$ symmetry 
\begin{align}
[\mm_0,E]=[\mm_0,F]=[\mm_0,G]=[\mm_0,\Gt]=0, 
\end{align}
where $\mm_0=\sum_{n=1}^{L-1}\mm_{n,n+1}$ 
is the Markov matrix for the closed ASEP without LK. 
\subsection{Steady state property}
Due to the $U_q(\mathfrak{sl}_2)$ symmetry described above, 
an $N$-particle stationary state $|S_N\ket$ without Langmuir kinetics ($\mm_0|S_N\ket=0$) 
is obtained by the successive actions of the creation operator $F$ on the vacuum $|0\ket$ 
together with the normalization constant $C_N$ as
\begin{align}
&|S_N\ket=C_NF^N|0\ket, \nn\\&
F^N|0\ket=
[N]_{q}!
\sum_{1\leq n_1<\cdots<n_N\leq L}q^{\sum_{j=1}^N(\nt_j-j)}|n_1, \cdots, n_N\ket, 
\end{align}
where $\nt_j$ is determined from $n_j$ through the relation $\nt_{N-j+1}=L-n_j+1$. 
This means that 
the $j$-th particle from the right is on the $\nt_j$-th site from the right 
if 
the $j$-th particle from the left  is on the $n_j$-th   site from the left. 
Here we introduce the $q$-factorial
\begin{align}
[N]_{q}!=[1]_q[2]_q\cdots[N]_q, \quad
[n]_q=1+q+\cdots+q^{n-1}. 
\end{align}
The constant $C_N$ is determined from the normalization condition
\begin{align}
\bra T_N|S_N\ket=1, 
\end{align}
where $|T_N\ket$ is the $N$-particle stationary state of the periodic ASEP without Langmuir kinetics 
\begin{align}
|T_N\ket:=\sum_{1\leq n_1<\cdots<n_N\leq L}|n_1, \cdots, n_N\ket. 
\end{align}
By use of the identity
\begin{align}
\sum_{1\leq n_1<\cdots<n_N\leq L}q^{\sum_{j=1}^N(n_j-j)}
=
\binom{L}{N}_{\!\!q}, 
\end{align}
we have
\begin{align}
|S_N\ket=&
\frac{[L-N]_q!}{[L]_q!}F^N|0\ket
=
\binom{L}{N}_{\!\!q}^{\!\!-1} 
\nn\\&
\times
\sum_{1\leq n_1<\cdots<n_N\leq L}q^{\sum_{j=1}^N(\nt_j-j)}|n_1, \cdots, n_N\ket, 
\label{sn}
\end{align}
where the $q$-binomial is defined by 
\begin{align}
\binom{L}{N}_{\!\!q}:=\frac{[L]_q!}{[N]_{q}![L-N]_q!}. 
\end{align}
For small $q$, the factor $q^{\sum_{j=1}^N(\nt_j-j)}$ represents the tendency for particles to gather to the right. 
In the case of $q=0$, the system is called a totally asymmetric simple exclusion process (TASEP), 
where particles move only to the right. 
In this case, we have $\nt_j=j$ and $[n]_q=1$, which lead to 
\begin{align}
|S_N\ket=F^N|0\ket=|L-N+1,\cdots, L-1, L\ket, 
\end{align}
where $N$ particles are completely clogged at the right end of the system. 
The exact density profile in the $N$-particle stationary state $\rho_N(x)$ is also known \cite{SS1994} 
\begin{align}
&\rho_N(x):=
\sum_{\tau_1,\cdots,\tau_L}\tau_x\bra\tau_1,\cdots,\tau_L|S_N\ket
\nn\\&=
\binom{L}{N}_{\!\!q}^{\!\!-1} 
\sum_{k=0}^{N-1}(-1)^{N-k+1}q^{\(N-k\)\(L-x-\frac{N+k-1}2\)}\binom{L}{k}_{\!\!q}. 
\label{rhoN}
\end{align}
\section{Periodic ASEP with Langmuir kinetics}
In the periodic case with Langmuir kinetics, the similarity transformation 
$A\to\tilde{A}=U^{-1}AU$
induced by the matrix
\begin{align}
U:=
\begin{pmatrix}1&1\\ \a&-1\end{pmatrix}
^{\otimes L}
\end{align}
is proved to be useful for the diagonalization of the corresponding Markov matrix \cite{SN}. 
Here $\a$ is the ratio between attachment $\om_a$ and detachment $\om_d$: 
\begin{align}
\a:=\frac{\om_a}{\om_d}. 
\end{align}
The stationary state is simply obtained by the operation of $U$ on the vacuum
\begin{align}
\mm_{\text{periodic}}\(U\vac\)=0, 
\end{align}
which is explicitly written as
\begin{align}
U\vac
=\binom1\a^{\otimes L}
=\sum_{N=0}^L\a^N\tn. 
\end{align}
The low-lying excitations are similarly constructed as
\begin{align}
\mm_{\text{periodic}}
|m\ket_{\text{periodic}}
=-m\om
|m\ket_{\text{periodic}}, 
\end{align}
where
\begin{align}
|m\ket_{\text{periodic}}:=U|T_m\ket
\quad
\text{for}
\quad
m=0,1,2,\cdots,L. 
\label{pbc_ex}
\end{align}
Noting that the relation $U^2=(1+\a)^L$, $\bra T|U=(1+\a)^L\bra0|$, 
we obtain the formula for the time evolution of a physical quantity $A$ 
starting from the vacuum initial state \cite{SN}
\begin{align}
\bra A(t) \ket
:=
\bra T|Ae^{\mm t}\vac
=\sum_{m=0}^L
e^{-m\om t}\a^m \bra 0|\tilde{A}|T_m\ket. 
\label{formula}
\end{align}
\section{Closed ASEP with infinitesimal Langmuir kinetics}
\subsection{Stationary state}
Now we consider the main object of this work: 
the closed ASEP with Langmuir kinetics. 
In this case the $\uq$ symmetry is broken due to the LK. 
The stationary state $|S\ket$ ($\mm|S\ket=0$) can no longer be written in a closed form 
and shows a very complicated one involving higher-order terms of $\om_a$ and $\om_d$. 
However, our finding is that the significant simplification occurs in the limit 
$\om:=\om_a+\om_d \to 0$ 
while keeping the ratio $\a=\om_a/\om_d$ finite. 
We propose a formula for the stationary state in this limit 
\begin{align}
\lim_{\om\to0}|S\ket
=
\frac1{(1+\a)^{L}}\sum_{N=0}^L
\binom{L}{N}
\a^{N}|S_N\ket, 
\end{align}
where $\sn$ is the $N$-particle stationary state \eqref{sn} in the case without Langmuir kinetics. 
We have not accomplished the proof of this formula yet. 
However, we confirm the result from direct analytical diagonalizations of the Markov matrix up to $L=4$. 

\begin{figure}
\includegraphics[width=0.9\columnwidth]{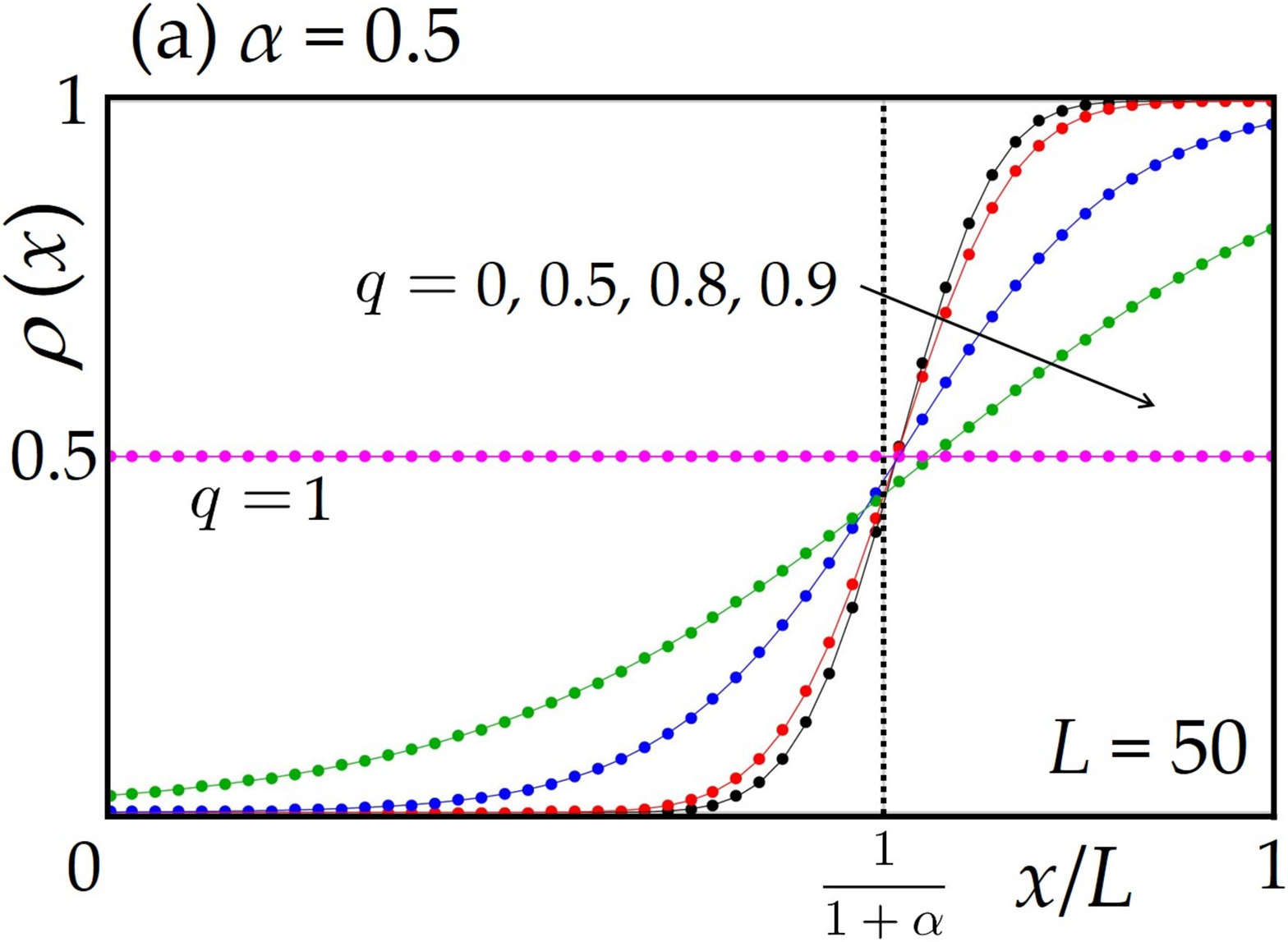}
\includegraphics[width=0.9\columnwidth]{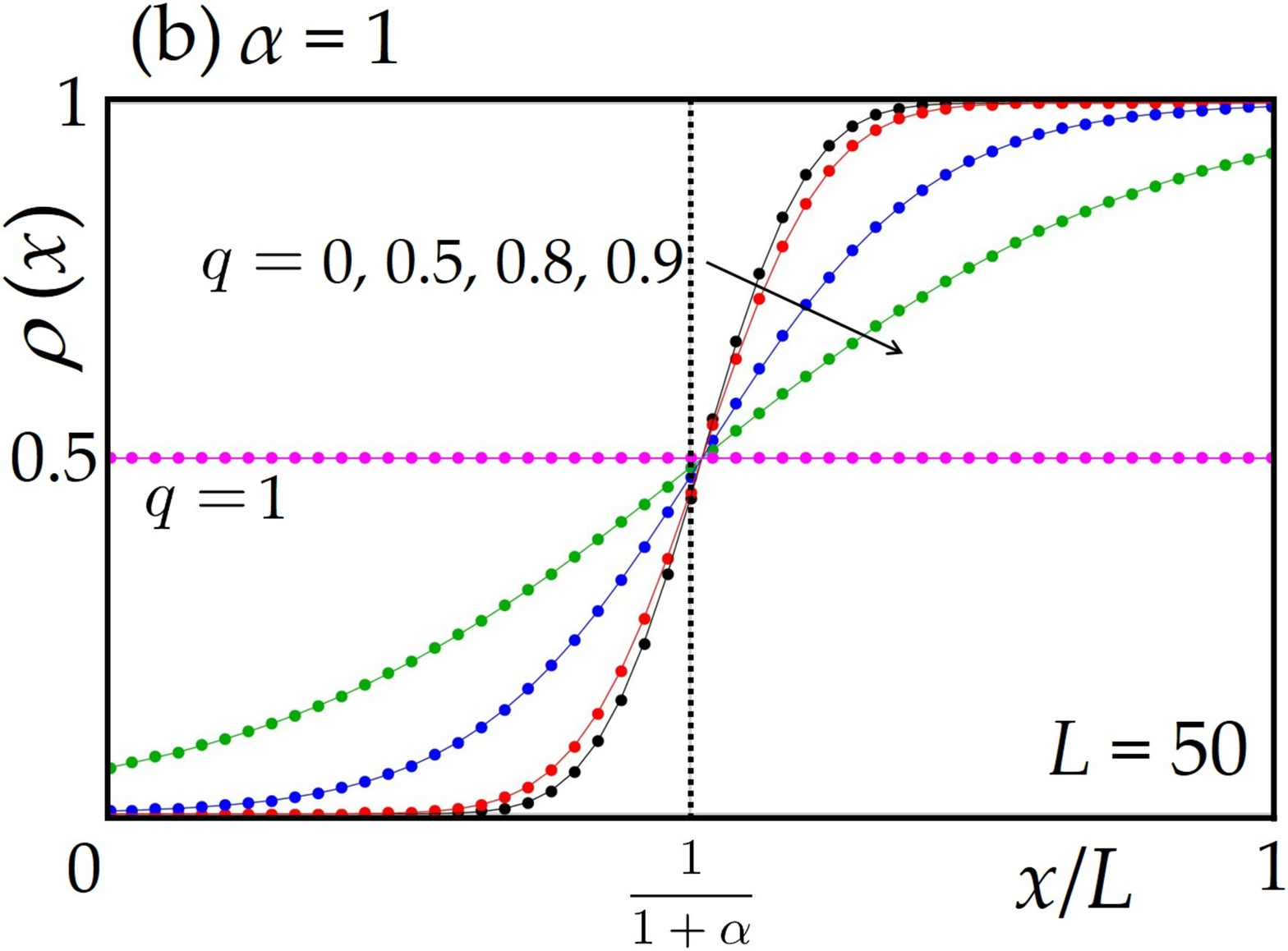}
\includegraphics[width=0.9\columnwidth]{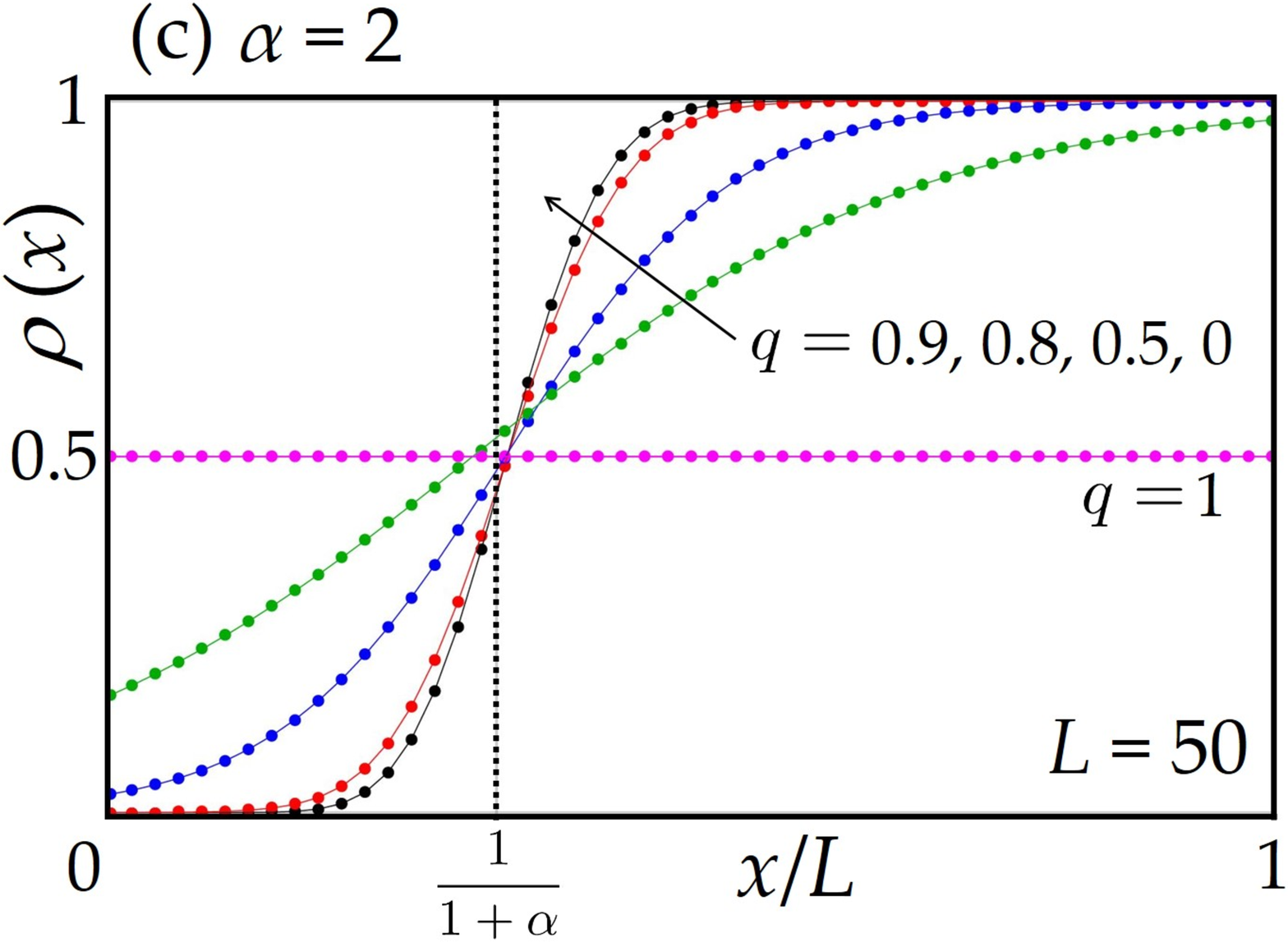}
\caption{ 
Density profiles in the steady state for $L=50$. 
Particles are gathered to the right due to the asymmetry of the hopping $q<p=1$. 
The end of the congestion is given by $x=L-\bra N \ket=L/(1+\a)$. 
In the case of symmetric diffusion ($p=q=1$), 
the profile becomes flat and translationally invariant. 
}
\label{tasep_density}
\end{figure}

By use of this formula, we immediately obtain the density profile $\rho(x)$ 
in the stationary state in the form
\begin{align}
\rho(x)&:=
\lim_{\om\to0}
\sum_{\tau_1,\cdots,\tau_L}\tau_x\bra\tau_1,\cdots,\tau_L|S\ket
\nn\\&=
\frac1{(1+\a)^L}
\sum_{N=0}^L
\frac{
\binom{L}{N}
}{
\binom{L}{N}_{\!\!q}
}
\a^N
\sum_{k=0}^{N-1}(-1)^{N-k+1}
\nn\\&\times
q^{\(N-k\)\(L-x-\frac{N+k-1}2\)}\binom{L}{k}_{\!\!q}. 
\end{align}
In Fig. \ref{tasep_density} 
we plot this formula for $\a=0.5, 1, 2$; $q=0, 0.5, 0.8, 0.9, 1$; and $L=50$. 

The expectation value of the particle number $N$ is calculated as
\begin{align}
\bra N \ket
=
\sum_{x=1}^L\rho(x)
=\frac{\a L}{1+\a}, 
\label{isotherm}
\end{align}
which coincides with the value known as the Langmuir isotherm \cite{Fowler}. 

\begin{figure}
\includegraphics[width=0.95\columnwidth]{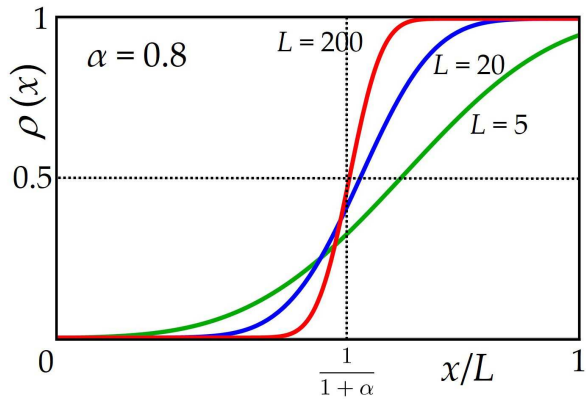}
\caption{ 
Density profiles in the steady state in the TASEP case ($q=0$)
for $L=5, 20, 200$. 
In this case the density profile is written in terms of the hypergeometric function \eqref{hg}. 
In the thermodynamic limit $L\to\infty$, the curve approaches the step function 
at the position determined from the Langmuir isotherm $\a/(1+\a)-1=1/(1+\a)$. 
}
\label{hyper_geom}
\end{figure}

In particular in the TASEP case, this formula can be further simplified into
\begin{align}
\rho(x)
&=
\frac1{(1+\a)^L}\sum_{N=y}^L\binom{L}{N}\a^N
\nn\\&
=
\binom{L}{y}\a^{y} F(L+1, y, y+1;-\a). 
\label{hg}
\end{align}
Here $y=L-x+1$ and 
$F(a,b,c;z)$ is the hypergeometric function
\begin{align}
F(a,b,c;z)
=
\sum_{n=0}^\infty
\frac{(a)_n(b)_n}{(c)_n n!}z^n, 
\end{align}
where
\begin{align}
(a)_n=a(a+1)\cdots(a+n-1).
\end{align}
In Fig. \ref{hyper_geom} 
we plot the density profile \eqref{hg} for $\a=0.8$ and $L=5, 20, 200$. 
In the case without Langmuir kinetics, the particle number $N$ is conserved and 
the density profile in the steady state gives the step function. 
On the other hand, in the case with Langmuir kinetics, 
the particle number $N$ fluctuates around the expectation value $\bra N \ket$ 
with the width $\D N\sim\sqrt{L}$. 

\subsection{Dynamics}
We also find the low-lying excitations
\begin{align}
\mm|m\ket=-m\om|m\ket, 
\label{ev}
\end{align}
which have the following closed form
in the limit $\om \to 0$ while keeping the ratio $\a$ finite:
\begin{align}
\lim_{\om\to0}
|m\ket=\sum_{N=0}^L\sn\bra T_N|U|T_m\ket
\quad
\text{for}
\quad
m=0,1,2,\cdots,L. 
\end{align}
These states are obtained by multiplying the excitations in the periodic case \eqref{pbc_ex} 
by the projection operator 
\begin{align}
P:=\sum_{N=0}^L\sn\bra T_N|
\end{align}
 from the left, 
which projects arbitrary $N$-particle states onto 
the $N$-particle stationary state $\sn$ \eqref{sn} without Langmuir kinetics. 
As in the case with the ground state, despite a lack of proof, 
this result is confirmed by 
direct analytical diagonalizations of the Markov matrix up to $L=4$. 

As in the periodic case \cite{SN}, 
these excitations are enough to analyze the dynamics from the vacuum initial state $\vac$. 
The time evolution of a physical quantity $A$ 
starting from the vacuum initial state $\vac$ is calculated as
\begin{align}
\bra A(t) \ket
:&=
\bra T|Ae^{\mm t}\vac
\nn\db&=
\frac1{(1+\a)^L}
\bra T|Ae^{\mm t}UU\vac
\nn\db&=
\frac1{(1+\a)^L}
\sum_{m=0}^L\a^m
\bra T|Ae^{\mm t}U\tm
\end{align}
The key idea is to 
insert the projection operator $P$, 
\begin{align}
\bra A(t) \ket
&=
\frac1{(1+\a)^L}
\sum_{m=0}^L\a^m
\bra T|Ae^{\mm t}
\nn\\&\times
\(
\sum_{N=0}^L
\sn\bra T_N|U\tm
\). 
\end{align}
The insertion of $P$ as an identity is justified in the limit $\om\to0$, 
where the relaxation time of the hopping process is much shorter than 
that of Langmuir kinetics $\om^{-1}$. 
Namely, an arbitrary $N$-particle state $|N\ket$ relaxes to the steady state 
$\sn\(=P|N\ket\)$ much faster than Langmuir kinetics. 
By use of the eigenvalue equation \eqref{ev}, we finally obtain the main formula of this work, 
\begin{align}
\bra A(t) \ket
&=
\frac1{(1+\a)^L}
\sum_{m=0}^L \a^m e^{-m\om t}
\nn\\&
\times
\binom{L}{m}
\sum_{\ell=0}^m
(-1)^\ell
\binom{m}\ell
\sum_{N=0}^{L-m}
\a^N
\binom{L-m}{N}
\bra A \ket_{N+\ell}
\label{formula}
\end{align}
where $\bra A \ket_{N}$ is the expectation value of $A$ 
in the $N$-particle steady state without Langmuir kinetics $|S_N\ket$, 
\begin{align}
\bra A \ket_{N}
:=\bra T_N|A|S_N\ket. 
\end{align}
Below we examine this formula for particle number and density profile as two simple examples. 
\subsubsection{Particle number}
\begin{figure}
\vspace{5mm}
\includegraphics[width=0.9\columnwidth]{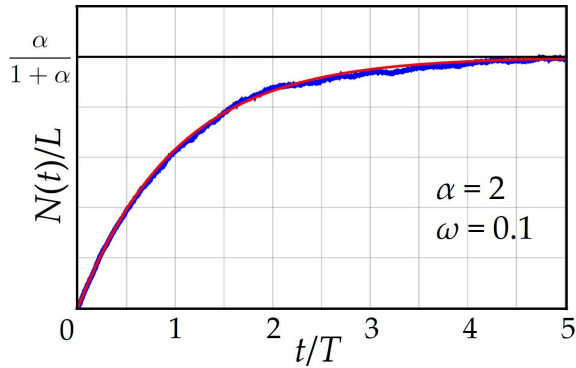}
\caption{ 
Time evolution of the number of particles $N(t)$ 
starting from the empty state
plotted against the scaled time $t/T$, 
where $T$ is the relaxation time $\om^{-1}$. 
The parameters are $\om=0.1$ and $\a=2.0$. 
The thick red line is the plot of our exact formula \eqref{density_formula}. 
The blue line is a 100-times average of Monte Carlo simulations. 
The parameters are $(p,q)=(1.0,0.3)$ and $L=100$ in the simulation. 
The plot starts from zero and saturates to the value $\a/(1+\a)$ known as the Langmuir isotherm. 
}
\label{particle_number}
\end{figure}

Let $A$ be a particle number operator $N=\sum_{n=1}^L\tau_n$. 
In this case, obviously, we have $\bra A \ket_{N}=N$. 
Substituting this into the formula \eqref{formula} gives 
\begin{align}
N(t)=\frac{\a L}{1+\a}\(1-e^{-\om t}\), 
\label{density_formula}
\end{align}
which represents an ordinary exponential relaxation 
to the Langmuir isotherm \eqref{isotherm} 
with a relaxation time $\om^{-1}$. 
Note that the terms with $m \geq 2$ are vanishing in \eqref{formula}. 
We observe good agreement with Monte Carlo simulations 
as shown in Fig. \ref{particle_number}. 

\subsubsection{Density Profile}
\begin{figure}
\includegraphics[width=0.8\columnwidth]{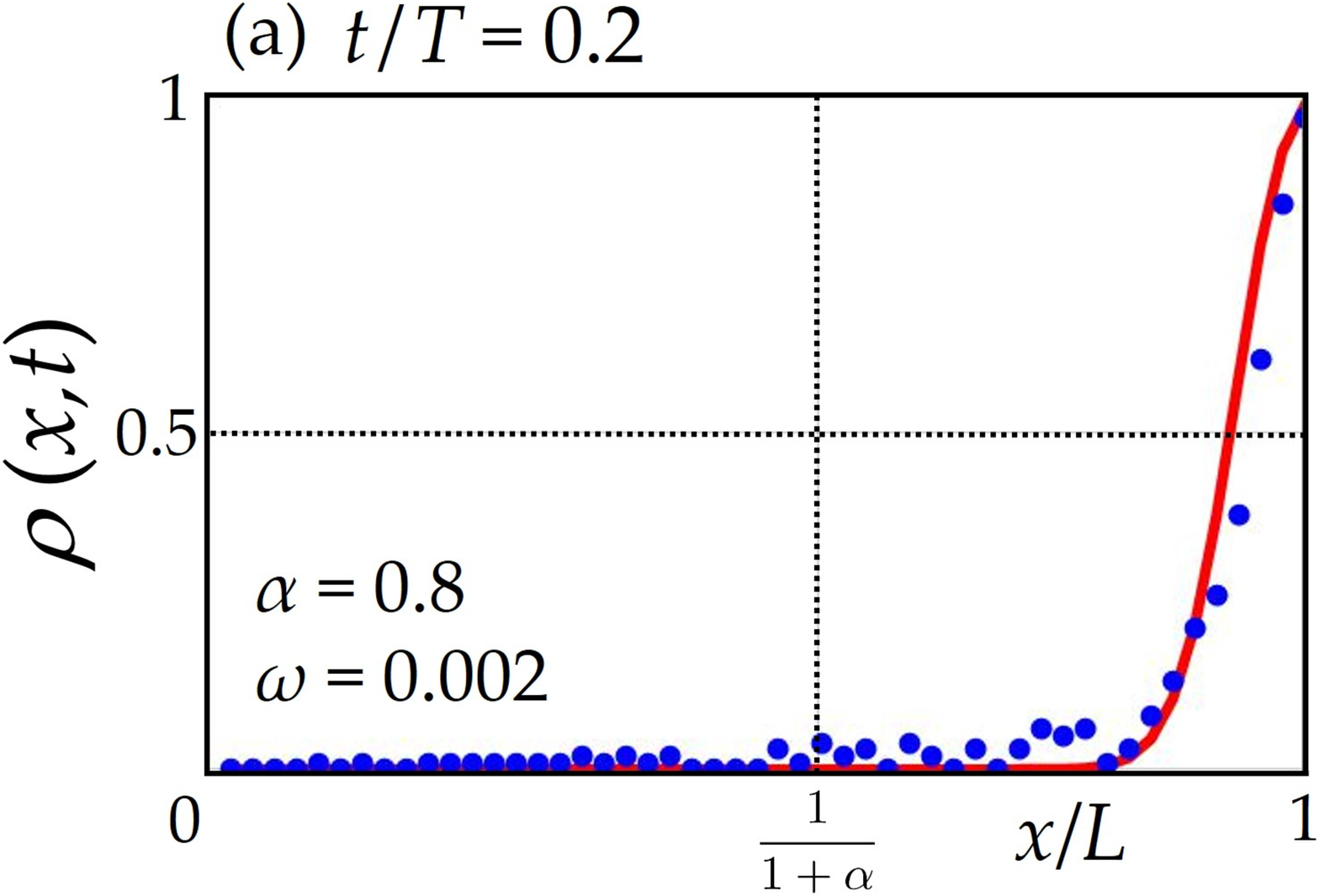}
\includegraphics[width=0.8\columnwidth]{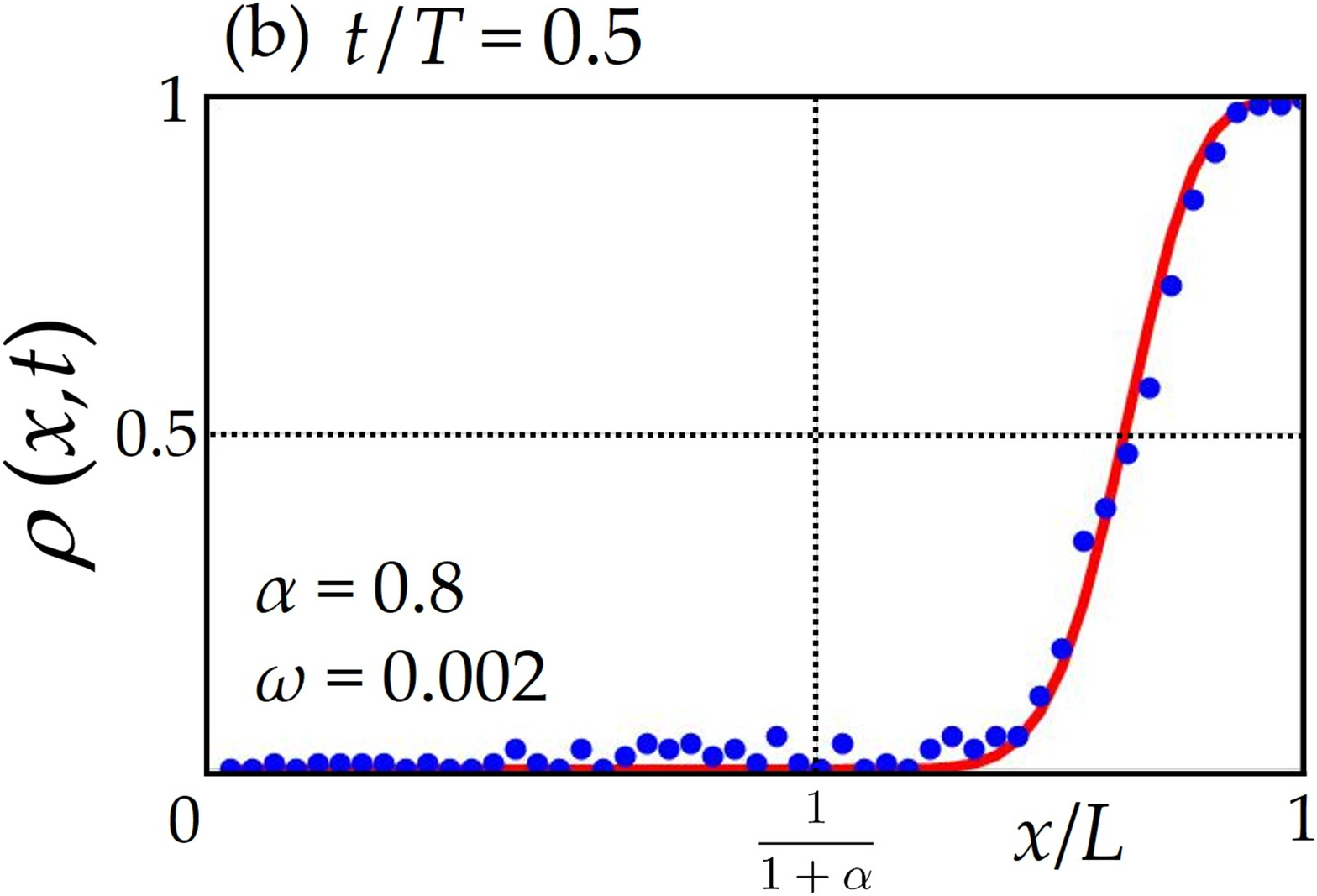}
\includegraphics[width=0.8\columnwidth]{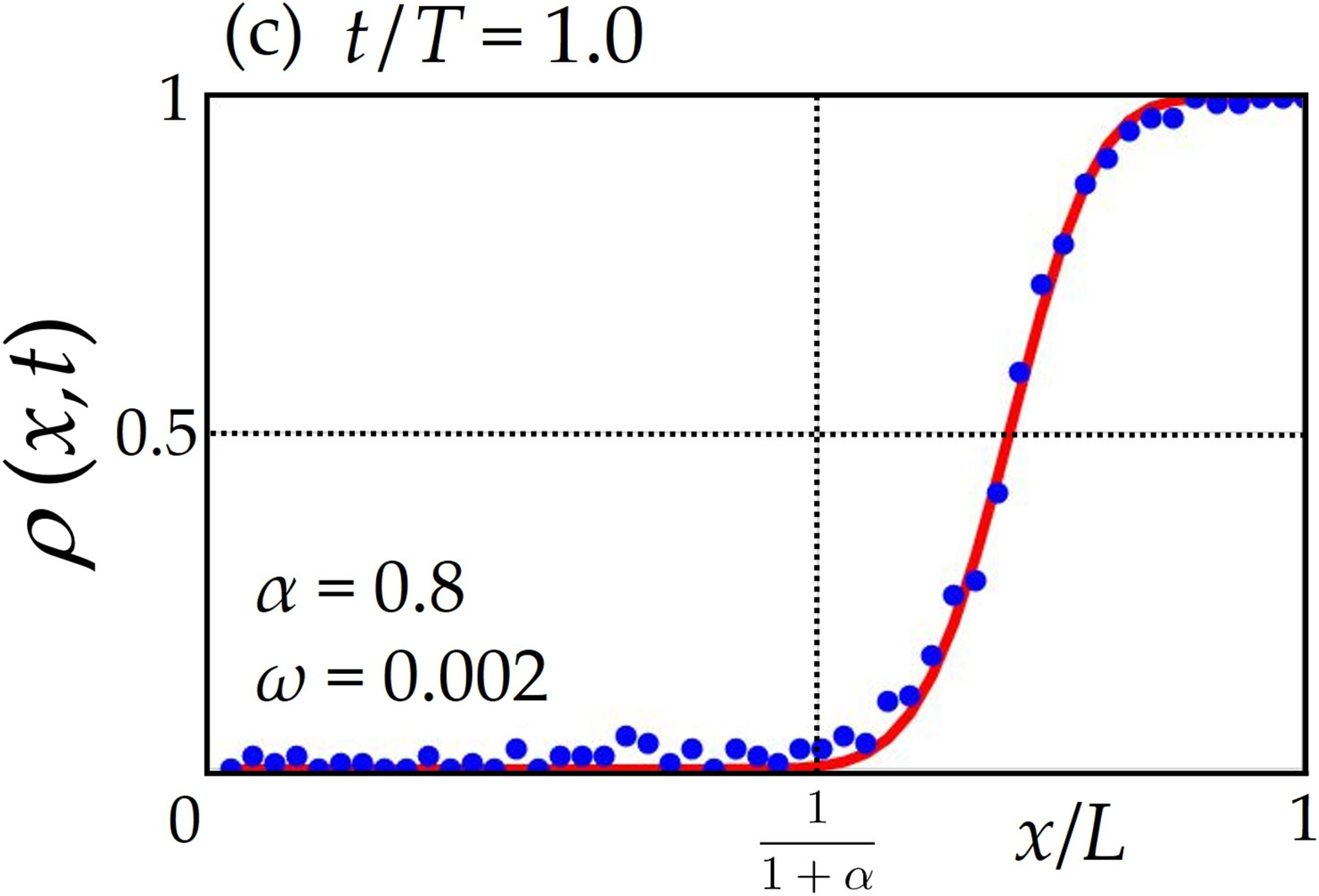}
\includegraphics[width=0.8\columnwidth]{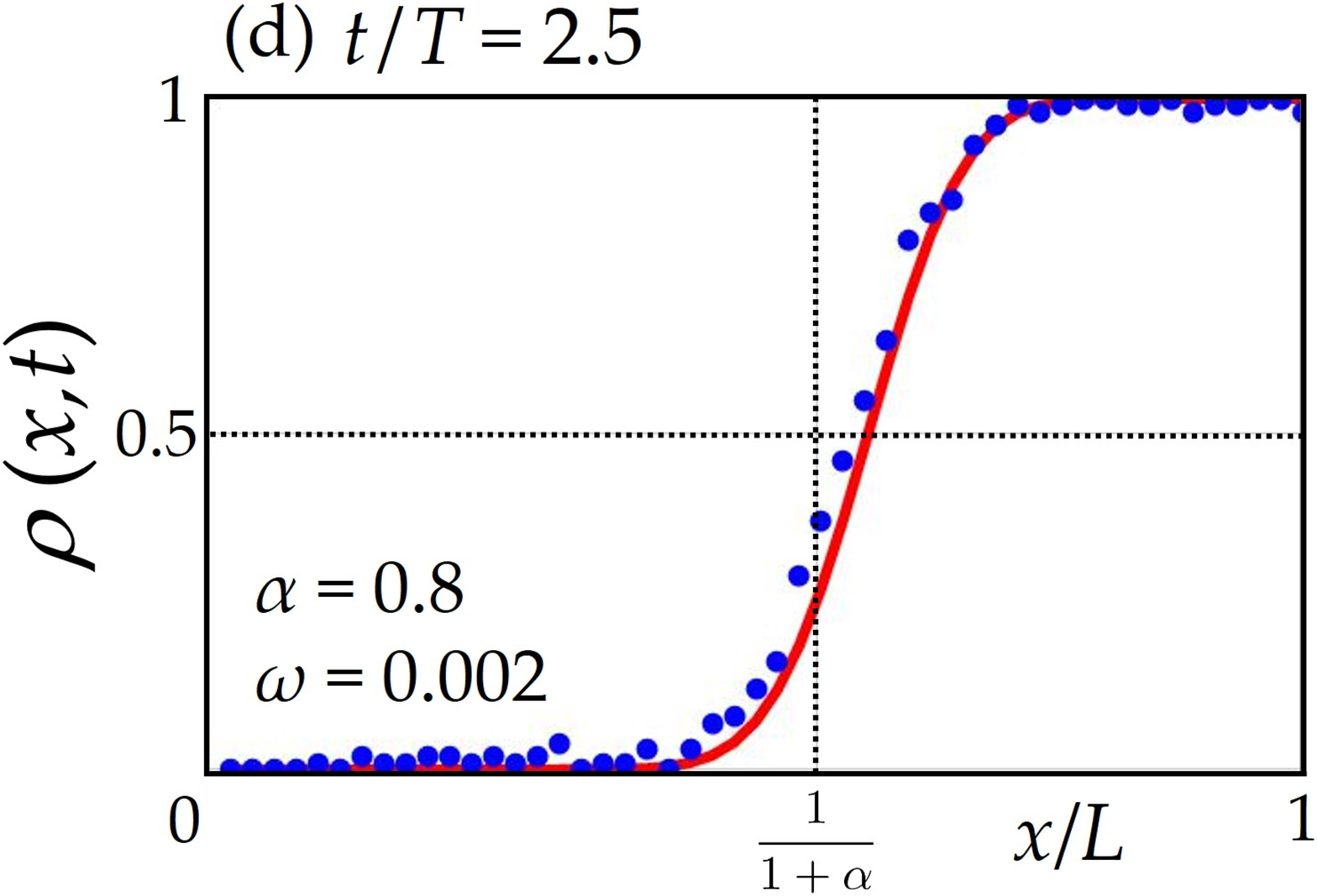}
\includegraphics[width=0.8\columnwidth]{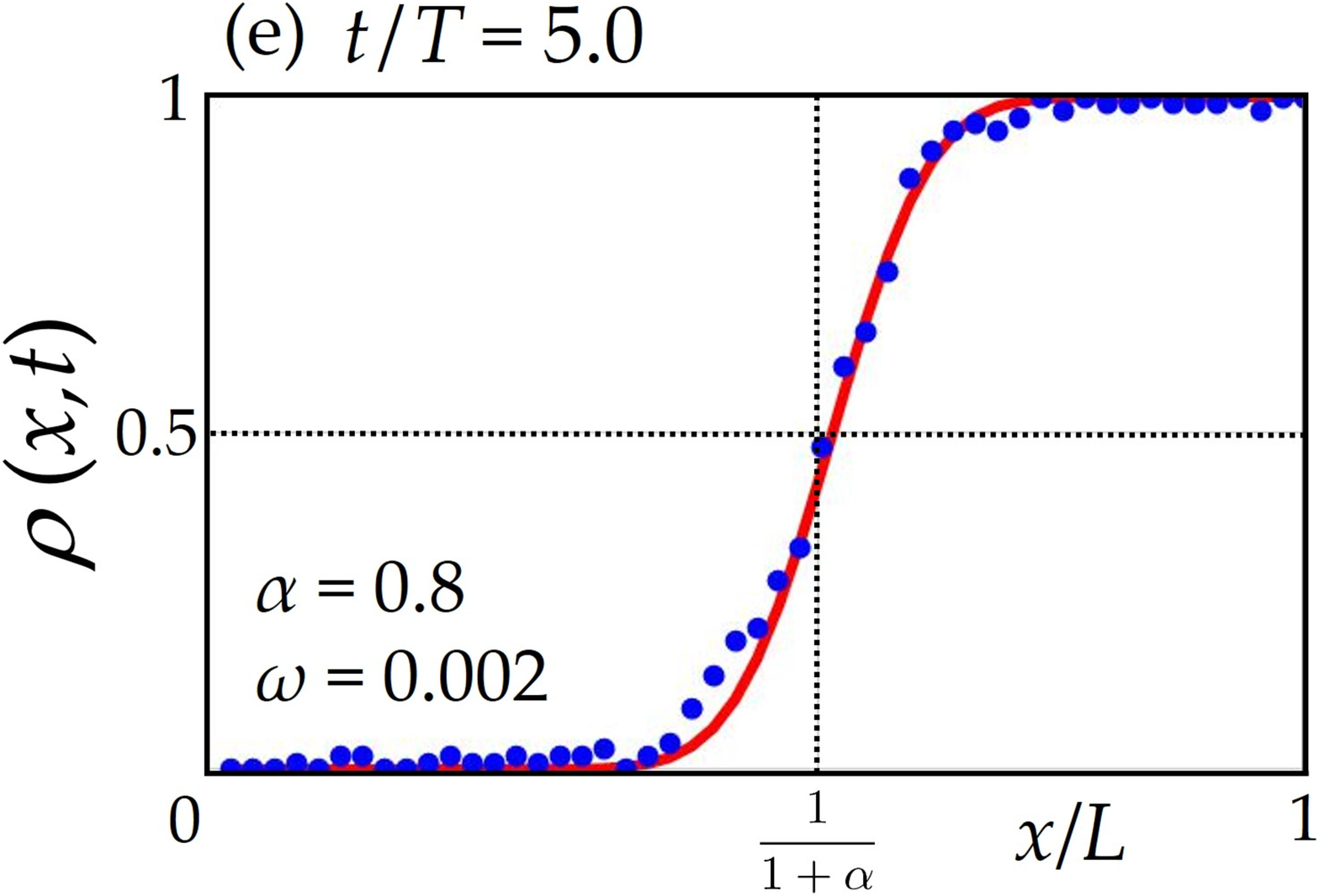}
\caption{ 
Time-dependent density profile $\rho(x,t)$. 
The time $t$ is scaled by the relaxation time $T=\om^{-1}$. 
The parameters are $(p,q)=(1.0,0.0)$, $\om=0.002$ and $\a=0.8$. 
The thick red lines are the plots of our exact formula \eqref{density_t_formula}. 
The blue dots are the result of 
a 100-time average of Monte Carlo simulations. 
We set the lattice length to $L=50$. 
The plot starts from zero and saturates to the curve 
described by the hypergeometric function \eqref{hg}. 
}
\label{dens_t}
\end{figure}

Next let us consider the time-dependent density profile $\rho(x,t)$. 
Substitution of $A=\tau_x$ into the formula \eqref{formula} yields
\begin{align}
\rho(x,t)
&=
\frac1{(1+\a)^L}
\sum_{m=0}^L \a^m e^{-m\om t}
\nn\\&
\times
\binom{L}{m}
\sum_{\ell=0}^m
(-1)^\ell
\binom{m}\ell
\sum_{N=0}^{L-m}
\a^N
\binom{L-m}{N}
\rho_{N+\ell}(x), 
\label{density_t_formula}
\end{align}
where $\rho_N(x)$ is the density profile in the $N$-particle steady state 
\eqref{rhoN}. 
In Fig. \ref{dens_t} 
we plot this formula \eqref{density_t_formula} together with the result of Monte Carlo simulations. 
We observe again good agreement with the simulations. 
\section{Conclusion}
In this paper we considered the asymmetric simple exclusion process 
in the closed boundary condition, 
with infinitesimally small Langmuir kinetics. 
In this limit, we conjectured the analytical form of the stationary state 
and a series of low-lying excited states. 
The correctness of these formula 
is supported by 
the analytical diagonalizations of the Markov matrix, 
although they lack a rigorous proof. 
Using this steady state and the excited states, 
we further proposed a formula for the time evolution of physical quantities 
starting from the empty state. 
As two simple examples, we computed a full time dependence of the number of particles and 
a density profile of the system. 
We observed good agreement with the Monte Carlo simulations in both cases. 
\section{acknowledgments}
The authors thank S. Ichiki for useful discussions. 


\begin{thebibliography}{n}
\bibitem{MGPB} 
C.T. MacDonald,  J.H. Gibbs and A.C. Pipkin, Biopolymers, {\bf 6} 1 (1968). 

\bibitem{DEHP} 
B. Derrida, M.R. Evans, V. Hakim and V. Pasquier J. Phys. A: Math. Gen. {\bf 26} 1493 (1993). 

\bibitem{KSKS} 
A.B Kolomeisky,  G.M. Sch$\rm{\ddot{u}}$tz, E.B. Kolomeisky and J.P. Straley, 
J. Phys. A: Math. Gen. {\bf 31} 6911 (1998). 

\bibitem{Sasamoto} 
T. Sasamoto, J. Phys. A: Math. Gen. {\bf 32} 7109 (1999). 

\bibitem{Schadschneider} 
A. Schadschneider, 
Physica A {\bf 285} 101 (2000).

\bibitem{SCN} 
A. Schadschneider, D. Chowdhury and K. Nishinari, 
{\it Stochastic Transport in Complex Systems: From Molecules to Vehicles} (Elsevier Science, Amsterdam, 2010). 

\bibitem{Schutz2000} 
G.M. Sch\"{u}tz, {\it Exactly solvable models for many-body systems far from equilibrium}, 
Phase Transitions and Critical Phenomena {\bf 19}, C. Domb and J. L. Lebowitz eds. (2000).
51


\bibitem{GM} 
O. Golinelli and K. Mallick, J. Phys. A: Math. Gen. 
{\bf 37} 3321 (2004); 
{\bf 38} 1419 (2005); 
{\bf 39} 12679 (2006). 
\bibitem{BE} 
R.A. Blythe and M.R. Evans, J. Phys. A: Math. Theor. {\bf 40} R333 (2007). 
\bibitem{BECE} 
R.A. Blythe, M.R. Evans, F. Colaiori and F.H.L. Essler, 
J. Phys. A {\bf 33} 2313 (2000). 
\bibitem{Krug} 
J. Krug, Phys. Rev. Lett. {\bf 67} 1882 (1991).
\bibitem{kpz} 
T. Sasamoto, H. Spohn, 
Phys. Rev. Lett. {\bf 104} 230602 (2010)

\bibitem{random} 
T. Sasamoto, 
J. Stat. Mech.: Theor. Exp. P07007 (2007).
\bibitem{PFF} 
A. Parmeggiani, T. Franosch and E. Frey, 
Phys. Rev. Lett. {\bf 90} 086601 (2003); 
Phys. Rev. E {\bf 70} 046101 (2004). 

\bibitem{EJS} 
M.R. Evans, R. Juh$\acute{\textrm{a}}$sz and L. Santen, 
Phys. Rev. E {\bf 68} 026117 (2003). 

\bibitem{ISN1} 
S. Ichiki, J. Sato, K. Nishinari,
J. Phys. Soc. Jpn. {\bf 85} 044001 (2016). 

\bibitem{ISN2} 
S. Ichiki, J. Sato, K. Nishinari,
Euro. Phys. J. B {\bf 89}: 135 (2016). 

\bibitem{EN} 
T. Ezaki and K. Nishinari, 
J. Phys. A: Math. Theor. {\bf 45} 185002 (2012). 
%

\bibitem{SN} 
J. Sato and K. Nishinari, 
Phys. Rev. E {\bf 93} 042113 (2016). 

\bibitem{SS1994} 
S. Sandow and G. Sch\"{u}tz, Europhys. Lett. {\bf 26} 7 (1994). 


\bibitem{Fowler} 
R.H. Fowler, {\it Statistical Mechanics} (Cambridge University Press, Cambridge, 1936). 


\end{thebibliography}
\end{document}